\documentclass[aps,prl,twocolumn, superscriptaddress,floatfix]{revtex4-1}
\usepackage{graphicx,bm,amssymb,amsmath,dcolumn,hyperref}
\usepackage{multirow}
\usepackage{enumerate}
\usepackage{enumitem}
\usepackage{color}
\usepackage{subfigure}  
\usepackage{epstopdf}
\usepackage{bbm}
\usepackage{graphicx}
\usepackage{hyperref}
\usepackage{threeparttable}
\usepackage{amsthm}
\usepackage{mathtools}
\usepackage{color}
\usepackage{tikz}
\usepackage{float}
\usepackage{algorithm}  
\usepackage{algorithmicx}  
\usepackage{algpseudocode}  
\usepackage{diagbox}

\newcommand\Fig[1]{Fig.~\ref{fig:#1}}

\newcommand{\Xnp}{X_{\textsc{np}}}

\newcommand{\scrR}{\mathcal{R}}

\def\intv#1[#2..#3]{\llbracket #2\mathrel{{.}\,{.}}\nobreak#3\rrbracket}

\begin{document}

\date{\today}\title{Nested Closed Paths in Two-Dimensional Percolation}
\author{Yu-Feng Song}
\affiliation{Hefei National Laboratory for Physical Sciences at Microscale and Department of Modern Physics, University of Science and Technology of China, Hefei, Anhui 230026, China}
\affiliation{MinJiang Collaborative Center for Theoretical Physics, Department of Physics and Electronic Information Engineering, Minjiang University, Fuzhou, Fujian 350108, China}
\author{Xiao-Jun Tan}
\affiliation{Hefei National Laboratory for Physical Sciences at Microscale and Department of Modern Physics, University of Science and Technology of China, Hefei, Anhui 230026, China}
\affiliation{Cainiao Network, Hanzhou, Zhejiang 310013, China}
\author{Xin-Hang Zhang}
\affiliation{Cainiao Network, Hanzhou, Zhejiang 310013, China}
\author{Jesper Lykke Jacobsen}
\email{jesper.jacobsen@ens.fr}
\affiliation{Laboratoire de Physique de l'\'Ecole Normale Sup\'erieure, ENS, Universit\'e PSL, CNRS, Sorbonne Universit\'e, Universit\'e de Paris, F-75005 Paris, France}
\affiliation{Sorbonne Universit\'e, \'Ecole Normale Sup\'erieure, CNRS, Laboratoire de Physique (LPENS), F-75005 Paris, France} 
\affiliation{Universit\'e Paris Saclay, CNRS, CEA, Institut de Physique Th\'eorique, F-91191 Gif-sur-Yvette, France}
\affiliation{Institut des Hautes \'Etudes Scientifiques, Universit\'e Paris Saclay, CNRS, Le Bois-Marie, 35 route de Chartres, F-91440 Bures-sur-Yvette, France}
\author{Bernard Nienhuis}
\email{nienhuis@lorentz.leidenuniv.nl}
\affiliation{Delta Institute of Theoretical Physics, Instituut Lorentz, P.O. Box 9506, NL-2300 RA Leiden, The Netherlands}
\author{Youjin Deng}
\email{yjdeng@ustc.edu.cn}
\affiliation{Hefei National Laboratory for Physical Sciences at Microscale and Department of Modern Physics, University of Science and Technology of China, Hefei, Anhui 230026, China}
\affiliation{MinJiang Collaborative Center for Theoretical Physics, Department of Physics and Electronic Information Engineering, Minjiang University, Fuzhou, Fujian 350108, China}

\begin{abstract}
  For two-dimensional percolation on a domain with the topology of a
  disc, we introduce a nested-path operator (NP) and thus a continuous
  family of one-point functions
  $W_k \equiv \langle \scrR \cdot k^\ell \rangle $, where $\ell$ is the
  number of independent nested closed paths surrounding the center, $k$
  is a path fugacity, and $\scrR$ projects on configurations having a
  cluster connecting the center to the boundary.  At criticality, we
  observe a power-law scaling $W_k \sim L^{-\Xnp}$, with $L$ the linear
  system size, and we determine the exponent $\Xnp$ as a function of
  $k$.  On the basis of our numerical results, we conjecture an
  analytical formula,
  $\Xnp (k) = \frac{3}{4}\phi^2 -\frac{5}{48}\phi^2/
  (\phi^2-\frac{2}{3})$ where $k = 2 \cos(\pi \phi)$, which reproduces
  the exact results for $k=0,1$ and agrees with the high-precision
  estimate of $\Xnp$ for other $k$ values.  In addition, we observe that
  $W_2(L)=1$ for site percolation on the triangular lattice with any
  size $L$, and we prove this identity for all self-matching lattices.
\end{abstract}


\maketitle

\paragraph{Introduction ---} 
Percolation~\cite{BroadbentHarmmersley57,StaufferAharony1994,Grimmett1999,BollobasRiordan2006}
is a paradigmatic model in the field of phase transitions and 
critical phenomena and a central topic in probability theory.
It also finds important applications in various fields such as fluids 
in porous media~\cite{hunt2014percolation},
gelation~\cite{stauffer1982gelation} and epidemiology~\cite{Meyers2007}.
Bond percolation corresponds to the $Q \! \to \! 1$ limiting case 
in the Fortuin-Kasteleyn cluster representation of the $Q$-state Potts model~\cite{Potts,FK},
and provides a simple illustration of many important concepts for {the latter~\cite{FYWu}}. 
In two dimensions (2D), the algebraic use of
symmetries---lattice duality~\cite{KW41}, Yang-Baxter integrability~\cite{Lieb67,Baxter72} and  conformal invariance~\cite{BPZ84,FQS84}---has led
to a host of {exact results.}
Critical exponents are predicted by Coulomb-gas (CG) arguments~\cite{Nienhuis1987}, conformal
field theory~\cite{Cardy1987} and Stochastic Loewner Evolutions~\cite{LawlerSchrammWerner2001}, 
and have been proven rigorously for e.g. triangular-lattice site percolation~\cite{Smirnov2001}.

\begin{figure}[t]
  \setlength{\unitlength}{.9\columnwidth}\begin{picture}(1,0.5)\put(0,0){
  \includegraphics[width=\unitlength]{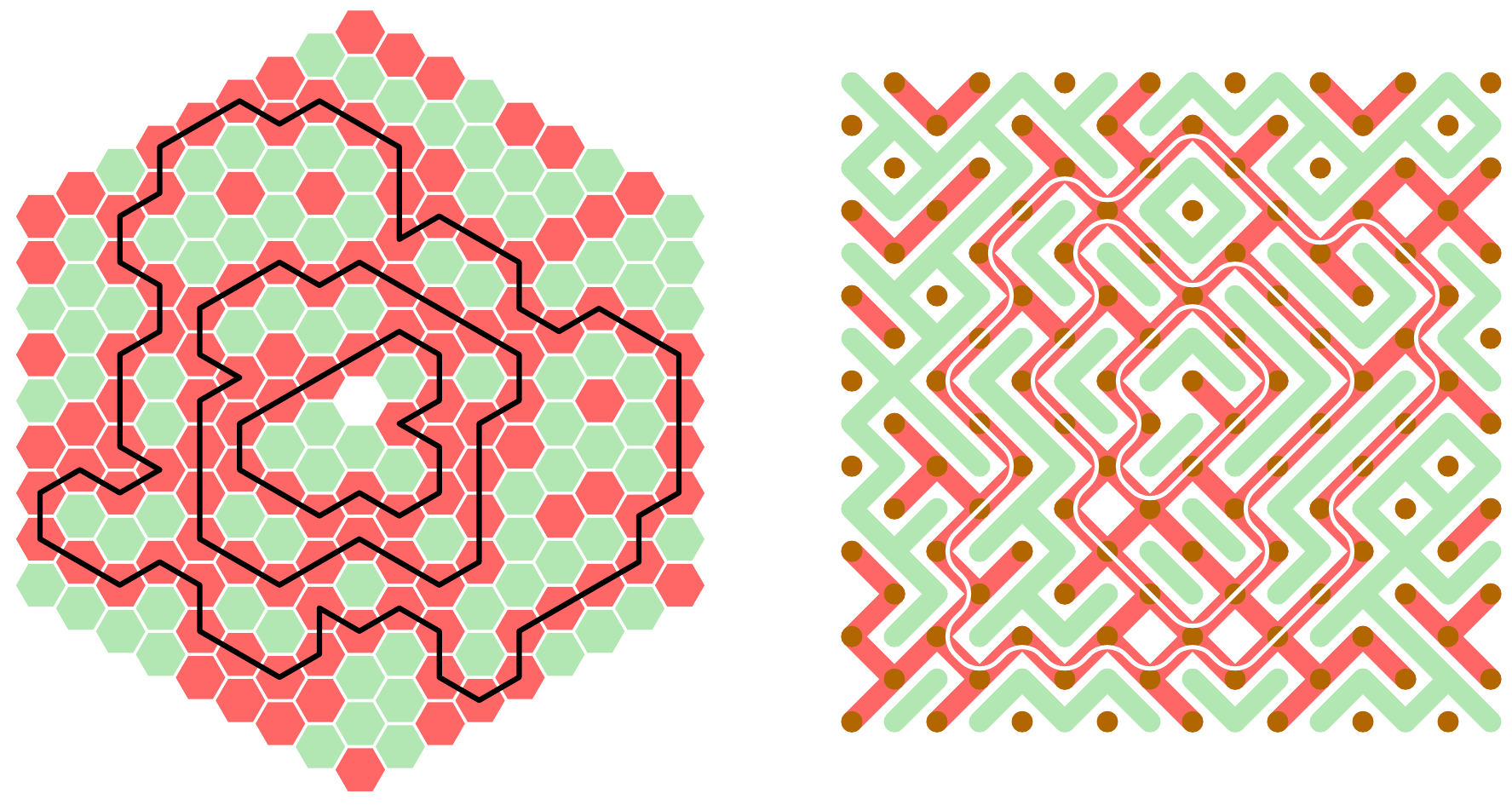}}\put(0.1,0){(a)}\put(0.75,0){(b)}\end{picture}
  \caption{Examples of configurations.
  (a) Triangular-lattice site percolation (STr).
  The central site is neutral (white), 
  the occupied (empty) sites are in red (green), 
  and the three independent nested paths are specified by black lines.
  (b) Square-lattice bond percolation (BSq). The occupied bonds
  are shown red and the unoccupied as green bonds on the dual
  lattice.  The neutral central bond is blank, and the three
  independent paths surrounding it are marked as white lines.
  They may visit the same site but not share a bond.
  }
  \vspace{-5mm}
 \label{fig:tris}
 \end{figure}

In site percolation, the sites of a lattice are occupied with probability $p$ and empty otherwise.
A sequence of distinct, occupied sites of which each is nearest neighbor to its predecessor is called a path.
Two occupied sites are connected if there is a path from one to the other. 
Two paths are independent if they do not have a site in common. 
A closed path has neighboring first and last sites. 
A maximal set of connected sites is called a cluster.

In bond percolation, the edges or bonds of the lattice are occupied with probability $p$, or vacant (empty).
Paths, connectivity and clusters follow the same definitions as in site percolation, 
with {\it sites} replaced by {\it bonds} of the lattice, and {\it nearest neighbor} 
by {\it having a site in common}.
Two paths are independent if they do not have a bond in common, and do not cross, 
but they may share a site.

Clusters and paths can also be introduced for empty elements. 
For bond percolation, these paths and clusters typically consist of bonds on the dual lattice. 
A dual bond is occupied if the (primal) bond it intersects is empty, 
and vice versa. For this reason, the paths on empty elements are often referred to as dual.

A path between two regions is typically not unique.  
In contrast, cluster boundaries form trajectories which are uniquely determined by the configuration.
Two different cluster boundaries are by nature non-overlapping.

Cluster boundaries define two families of exponents, which have been computed by CG methods.  
The so-called {\it watermelon} exponents~\cite{BN1984,DupSal1987} govern the probability 
that two (or more) distant regions are connected by a given number, say $n$, of cluster boundaries.  
The watermelon exponents have the value $X_{\textsc{wm}}(n) =(n^2-1)/12$.

The second, continuous family governs the correlation function of 
what we here here call the {\it nested-loop} operator (see \cite{dNijs1983, MitNie2004}). 
This operator gives a weight, say $k$, to each cluster boundary surrounding the insertion point 
(multiple such boundaries must be nested, whence the name). 
Its two-point function gives a weight to those cluster boundaries that surround one, 
but not both of the insertion points. The corresponding exponent is
\begin{equation} 
  X_{\textsc{nl}}({k}) =  \frac{3}{4}\phi^2 -  \frac{1}{12} \,,
  \quad\text{ where } {k} = 2 \cos(\pi\phi)\,. \label{eq:NL-exp}
\end{equation}

Analogous to $X_{\textsc{wm}}(n)$ are the so-called monochromatic $n$-path 
exponents $X_{\textsc{mp}}(n)$, governing the decay of probability 
that between two distant regions there are $n$ independent paths, 
all on clusters (or, equivalently, all on dual clusters). 
The case $n\!=\!1$ means that no cluster boundary separates the two regions, 
so $X_{\textsc{mp}}(1) \!=\! X_{\textsc{nl}}(0) \!=\! \frac{5}{48}$. 
But the exponents for $n=2$ (backbone exponent) or higher $n$ do not appear amenable to CG analysis, 
and hence they are known only numerically~\cite{JacZinn02,BefNol2011,Xu2014}.  
As a side-remark we mention that when one or more, but not all, of the paths are on dual clusters, 
the exponents are different and in fact identical to $X_{\textsc{wm}}(n)$ \cite{Aizenman1999, BefNol2011}.  

In this Letter we propose to similarly consider the path analogue of the nested-loop operator: 
the nested-path (NP) operator.  
It gives a weight $k$ to each independent closed path surrounding the insertion point. 
We investigate here the exponent of this operator by numerical means.  
For simplicity, we simulate its one-point function: 2D percolation with the NP operator 
placed at the center of a compact domain of linear dimension $L$.  
We thus estimate the expectation value $W_k\equiv\langle \scrR\cdot k^\ell \rangle$, 
where $\scrR$ is the indicator function of a path from the center to the  boundary of the domain, 
and $\ell$ the number of independent closed paths surrounding the center.  
The factor $\scrR$ ensures that two consecutive surrounding paths are connected, 
rather than separated by two cluster boundaries.  The configurations with $\scrR=1$ we call percolating.

\paragraph{Results, summary ---}
We show that, at the percolation threshold, the scaling of $W_k$ obeys a power law
$W_k \! \sim \!  L^{-\Xnp}$ 
with an exponent $\Xnp (k)$, that depends continuously on the weight $k$.
A high-precision estimate of $\Xnp$ is obtained as a function of $k$.
For $k=2$, we observe that $W_2(L)=1$ for site percolation 
on the triangular lattice with any $L$, and prove this to be true for any self-matching lattice~\cite{SykesEssam64}.
We present an analytical formula (\ref{eq:zeta_conj}), analogous to (\ref{eq:NL-exp}), 
which reproduces some exact values and agrees so well with the numerical results, 
that we conjecture it to be exact.

\begin{figure}[t]
  \includegraphics[width=1.0\columnwidth]{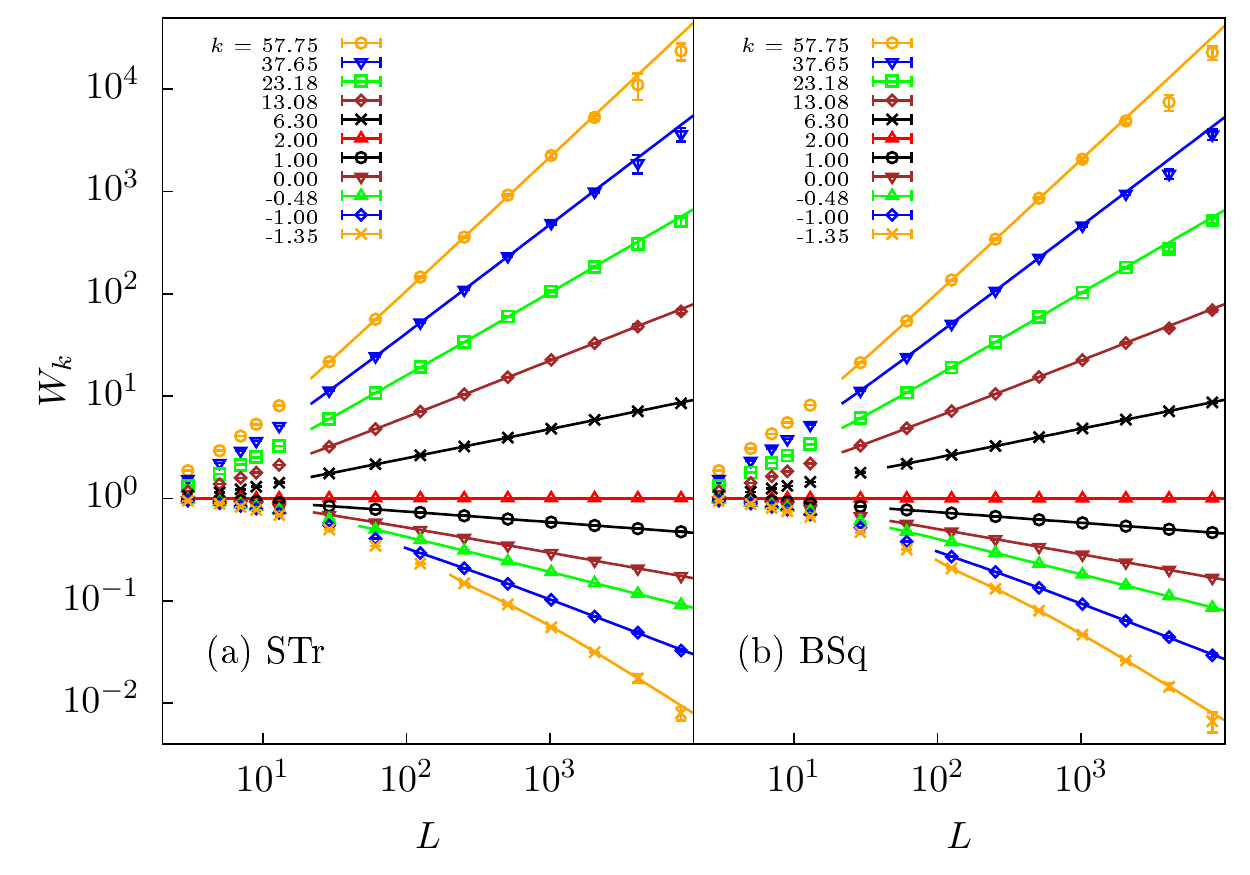}  \vspace{-4mm}
  \caption{Log-log plot of $W_k$ versus linear size $L$, for STr (a) and BSq (b).
  The lines represent the fitting curves by Eq.~(\ref{eq:fit}) and strongly indicate the algebraic 
  dependence of $W_k$ on $L$. }
  \vspace{-5mm}
  \label{fig:WkvsL}
\end{figure}

\paragraph{Results, details ---}
We study site percolation on a triangular lattice (STr) in a hexagonal domain with free boundaries.
The scale $L$ is the length of the diagonal.  
\Fig{tris}a shows a sample configuration, with $L=17$.
The central site is neutral, and the other sites are occupied with probability $p$.
For each configuration we calculate the number $\ell$ of {\it independent} closed paths that surround the center, 
and $\scrR$ which is 1 (0) if there is (not) a path from the center to the boundary.  
While $\ell$ is well-defined, the paths are not unique. In \Fig{tris}a, $\scrR=1$ and $\ell=3$.

Analogous procedures are applied to bond percolation on the square lattice (BSq), 
see \Fig{tris}b, with a neutral bond placed at the center, and the length of the diagonal, $L=15$.

We are interested in the scaling behavior of $W_k(L)$ at the percolation threshold $p_c$.
For $k=1$, $W_1 \equiv \langle \scrR  \rangle$ represents the probability 
that the central site is connected to the boundary, 
which is seen from \eqref{eq:NL-exp} to decay as $\langle \scrR  \rangle \sim L^{-5/48}$.
The contributions to $W_0$ are those, which have a path from the center to the boundary, 
but {\it no} closed path surrounding the center.
These configurations must have a path of occupied and 
one of empty elements from the center to the boundary 
and consequently two cluster boundaries connecting the center to the boundary.  
These events are selected for the watermelon exponent $X_{\textsc{wm}}(2) = 1/4$.
Thus, proposing $W_k \sim L^{-\Xnp(k)}$, we already know that $\Xnp(1)=5/48$ and $\Xnp(0)=1/4$. 

\begin{figure}[h]
\includegraphics[width=1.0\columnwidth]{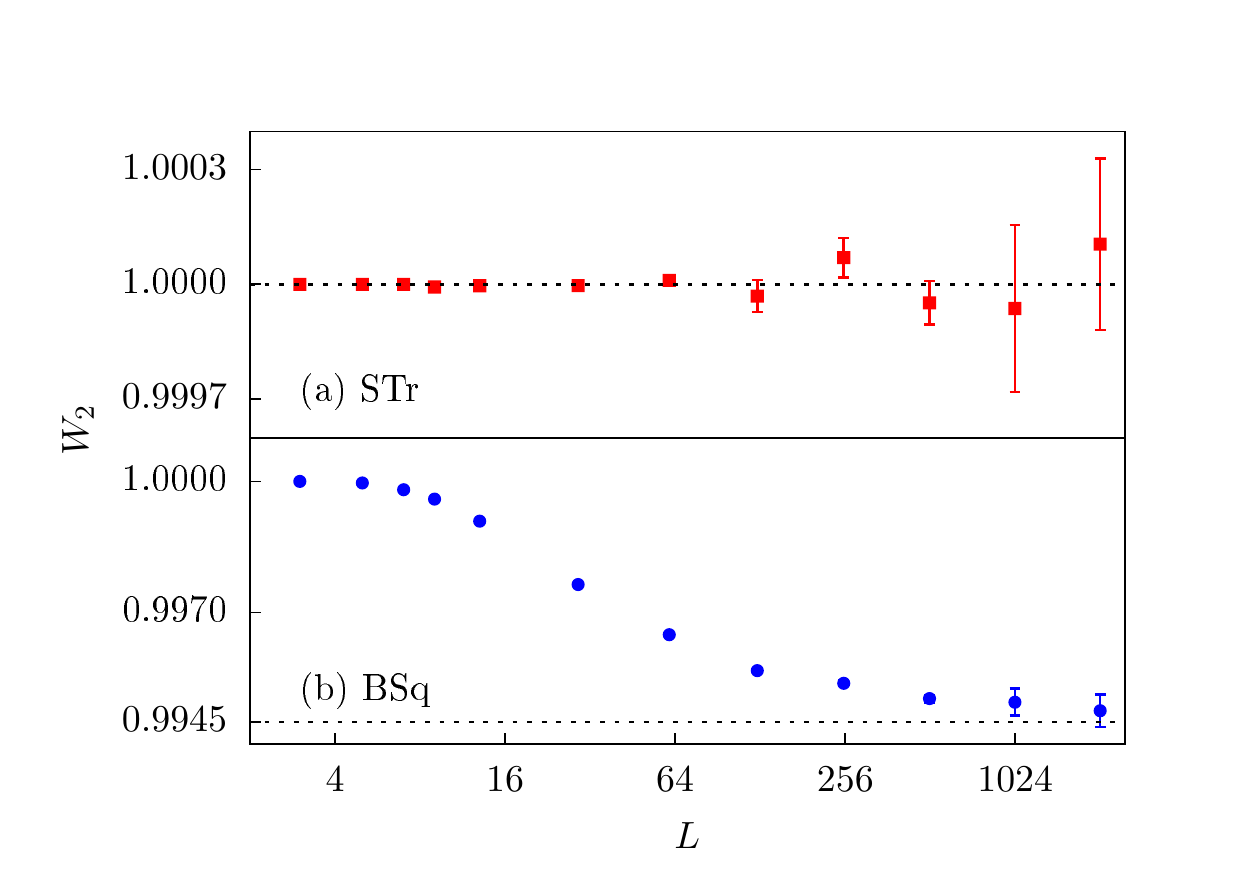} \vspace{-4mm}
\caption{Observable $W_2$ versus $L$, for STr and BSq.
The STr values for $L\leq 7$  are exact.
}
\vspace{-5mm}
  \label{fig:W2vsL}
\end{figure}
We carry out extensive  simulations for STr and BSq
at the percolation threshold $p_{\rm c}=1/2$, with geometric shapes as in \Fig{tris}. 
The linear size $L$ is taken in range $ 3 \leq L \leq 8189$.
For both models and each $L$, the number of samples is at least $5 \times 10^9$  
for $L \leq 100$, $2 \times 10^8$ for $100 < L \leq 1000$, 
$2 \times 10^7$ for $100 < L \leq 4000$, and $4 \times 10^5$ for $L >4000$.

A log-log plot of $W_k$ versus  $L$ is shown in \Fig{WkvsL}.  
The data clearly support asymptotic power-law dependence of $W_k$ on $L$.
We fit (by least squares) the $W_k$ data to
\begin{equation}
W_k = L^{-\Xnp} (a + b_1 L^{-\omega} + b_2 L^{-2 \omega}) \,.
\label{eq:fit}
\end{equation}
We admit only data points with $L \geq L_{\rm m}$ for the fits, and systematically study the effect
on the residual $\chi^2$ value (weighted according to confidence level) when varying $L_{\rm m}$. 
In the best fits, $\omega \approx 1$.
The results with $\omega =1$ are given in Tab.~\ref{tab:zeta}.
The estimates of $\Xnp(1)$ and $\Xnp(0)$ agree well with the exact values $5/48$ and $1/4$, respectively.
Table~\ref{tab:zeta} strongly indicates that  $\Xnp(2)=0$ for $k=2$. 

	\begin{table}[!ht]
  \begin{tabular}{c|ccccc}
	\hline
  \hline
  $\phi^2$           & -1          & -3/4        & -1/2        & -1/4	       & 0   \\
  $k$                & 23.18         & 15.26         &	9.329    & 5.018     & 2 \\
	\hline
	conj  & -0.8125\;\,    & -0.6177\;\;\,    & -0.41964\;\;\,    & -0.21591\,    & 0\hspace{33pt} \\
	STr    &\;-0.813(5)   & -0.619(2)   & -0.421(1)\;\,   & -0.216(1)   & -0.0000(2)\;   \\
	BSq    &\;-0.810(2)   & -0.617(1)   & -0.4192(5)  & -0.215(2)   & 0.0004(5)	  \\
	\hline 
  \hline
  $\phi^2$           & 1/9       & 1/4       & 1/3       & 3/8       & 4/9   \\
	$k$                & 1         & 0         & -0.4812     & -0.6915     &-1 \\
  \hline
  conj   & \;0.10417\;\;\;   & 0.25\hspace{21pt}      & 0.3542\;\;\;    & 0.4152\;\;\,    & 0.5417\;\;\,\\
	STr    & \;0.1043(2) & 0.2500(3) & 0.354(2)  & 0.414(2)  & 0.544(6)      \\
	BSq    & \;0.1044(4) & 0.2503(6) & 0.355(1)  & 0.416(2)  & 0.551(6)      \\
	\hline
	\hline
	\end{tabular}
	\caption{Fitting results of $\Xnp(k)$ as a function of $k$, with the rows `conj' 
  from Eq.~(\ref{eq:zeta_conj}).  
  }
	\label{tab:zeta}
	\end{table}

{The $W_2$ data for STr and BSq are listed in Tab.~\ref{tab:W2} and
plotted in \Fig{W2vsL} versus $L$.}
For STr we find for $L \leq 7$ by exact enumeration, 
and for larger $L$ within statistical errors, that $W_2=1$.  
{For BSq, we find $W_2=1$ for $L=3$ and $2097075/2^{21}$ for $L=5$, shown in  Tab.~\ref{tab:W2} together with simulation data for larger $L$.
As $L$ increases, $W_2$ converges to a value slightly smaller than, 
and clearly different from 1.}
A least-squares fit $W_2(L) = W_{2, \infty} + b L^{-2}$ for $L>20$, 
gives the asymptotic value as $W_{2,\infty}=0.994\,5(2)$.
Thus, we conjecture $\Xnp(2)=0$ in general, and, for STr, $W_2(L)=1$.

\begin{table}[t]
	\begin{tabular}{c||l|l|l|l}
		\hline
		\hline
		$L$ & \hspace{20pt}3 & \hspace{20pt}5 & \hspace{20pt}7 & \hspace{20pt}9 \\
		\hline
		STr 	& 1 & 1 & 1 & 0.999993(4)  \\
		BSq	  & 1 & 0.999963283 & 0.999807(5) & 0.999594(6) \\
		\hline
		\hline
		$L$  & \hspace{16pt}13  & \hspace{16pt}29 &\hspace{16pt} 61 & \hspace{16pt}125 \\
		\hline
		STr	 & 0.999997(6)  &0.999997(7) & 1.000010(9) & 0.99997(4) \\
		BSq  & 0.999088(5)  & 0.99764(1) & 0.99649(1) & 0.99567(5) \\
		\hline \hline
		$L$ &\hspace{11pt} 253 & \hspace{16pt}509 &\hspace{10pt} 1021  & \hspace{10pt}2045  \\
		\hline
		STr 	 & 1.00007(5) & 0.99995(6) & 0.9999(2)   & 1.0001(2)  \\ 
		BSq 	 & 0.99538(6) & 0.99503(8) & 0.9949(3)  & 0.9948(4) \\ 
		\hline
		\hline
	\end{tabular}
  \caption{Observable $W_{2}$.  
  For $L=4093$ and 8189,  we have respectively $W_2=0.999(1)$ and 0.999(2) for STr,
  and $W_2=0.995(2)$ and $0.994(2)$ for BSq.
  }
	\label{tab:W2}
\end{table}

  \begin{figure}[b]
    \includegraphics[width=1.0\columnwidth]{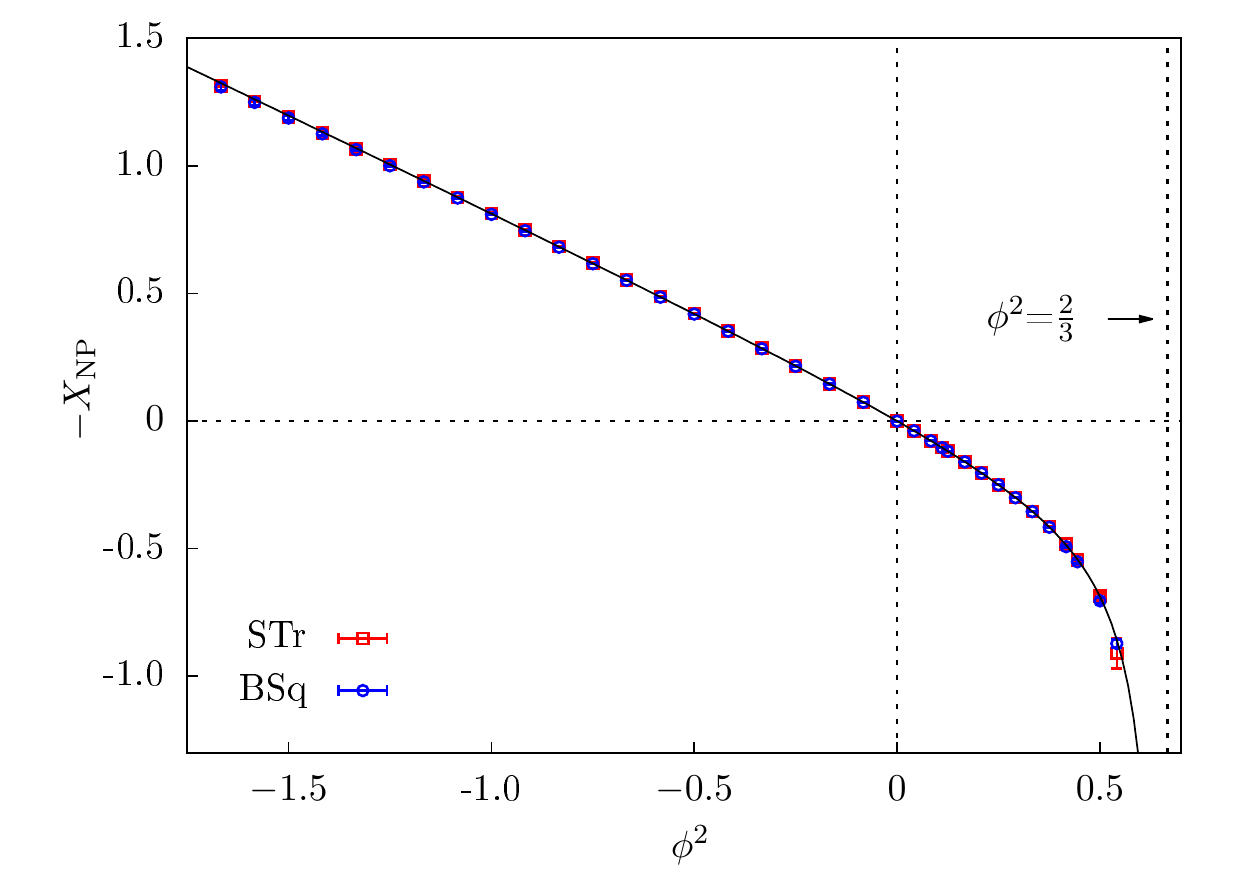} \vspace{-4mm}
    \caption{Exponent $\Xnp(k) $ versus $\phi^2$. 
    Estimates of $\Xnp$ for STr and BSq agree well with Eq.~(\ref{eq:zeta_conj}),
    shown as the solid curve.} 
    \vspace{-5mm}
    \label{fig:exp}
  \end{figure}

Then, in an attempt to find a formula analogous to (\ref{eq:NL-exp}),
we set $k=2 \cos (\pi \phi )$ and plot
$\Xnp$ as a function of $\phi^2$, as shown in \Fig{exp}.
Noting (i) $\Xnp(2)=0$,  (ii) an apparent pole for some positive $\phi^2<1$, 
and (iii) an asymptotic slope of 3/4 for negative $\phi^2$,  
just as $X_{\textsc{NL}}$, (\ref{eq:NL-exp}), it is natural to propose $\Xnp = 3\phi^2/4 + a\phi^2/(\phi^2-b)$ 
as the simplest rational function that matches these observations.  
Then the exact results for $\Xnp(0)$ and $\Xnp(1)$ fix
\begin{equation}
\Xnp(k) \;=\;  \frac{3}{4} \phi^2 - \frac{5}{48}\;\frac{\phi^2}{(\phi^2-2/3)},
\label{eq:zeta_conj}\end{equation}
in which some well-known exponents of 2D percolation seem to appear.
The excellent agreement with the numerical results 
shown in \Fig{exp} and Table~\ref{tab:zeta}, 
leads us to conjecture that (\ref{eq:zeta_conj}) is exact.  
But, in spite of some similarity with (\ref{eq:NL-exp}) 
we have not found any theore\-tical basis for (\ref{eq:zeta_conj}).

\paragraph{Method ---} In simulations, each site (bond) is randomly occupied with  probability $p_c=1/2$,
a cluster is grown from the center by standard breadth-first search.
If the cluster does not reach the boundary, ${\scrR} =0$ and there is no contribution to $W_k$.
Otherwise, we compute the number $\ell$ of paths surrounding the center.
This can be achieved efficiently by the following algorithm.

The $m$-th path surrounding the center acts as the seed of one or more dual (empty) clusters, linked together by the $(m-1)$-th surrounding path.
By growing the dual clusters from the $(m-1)$-th surrounding path, one 
can locate the $m$-th path as the chain of occupied sites (or bonds) that stops the dual-cluster growth.
A caveat is that for BSq the dual cluster consists of bonds of the {\em dual} lattice.
Algorithm~\ref{alg1} sketches the corresponding procedure, in which $\Omega$ is the region that is encircled by the next surrounding path.  It is written for STr, but when sites are replaced by bonds, and adjacent sites by bonds sharing a site, it works for BSq too.

\floatname{algorithm}{Algorithm}
\begin{algorithm}[H] 
	\caption{Calculate $\ell$ for site percolation}
\begin{algorithmic}
         \State $\ell =0$ 
         \State $\Omega \gets$ the central site
         \State Grow the empty cluster around $\Omega$
	\While{boundary is not reached}
	\State $\ell = \ell+1$
	\State $\Omega \gets$ set of occupied sites encircling the empty cluster
	\State Grow the empty cluster around $\Omega$
	\EndWhile
\end{algorithmic}\label{alg1}
\end{algorithm}

\paragraph{Proof of the identity} $W_{2}(L)=1$ for {\rm STr} --- 
Take an arbitrary STr configuration on a simply connected piece of the lattice, 
with a distinguished, `central' site. 
Consider a maximal set of independent paths surrounding the center, with each path consisting either of only occupied (red) sites or of only empty (green) sites.
The {\em innermost} such colored paths, given the interior ones, are uniquely defined 
(and can be constructed by Algorithm~\ref{alg1}).

\begin{figure}[t]
 \includegraphics[width=1.0\columnwidth]{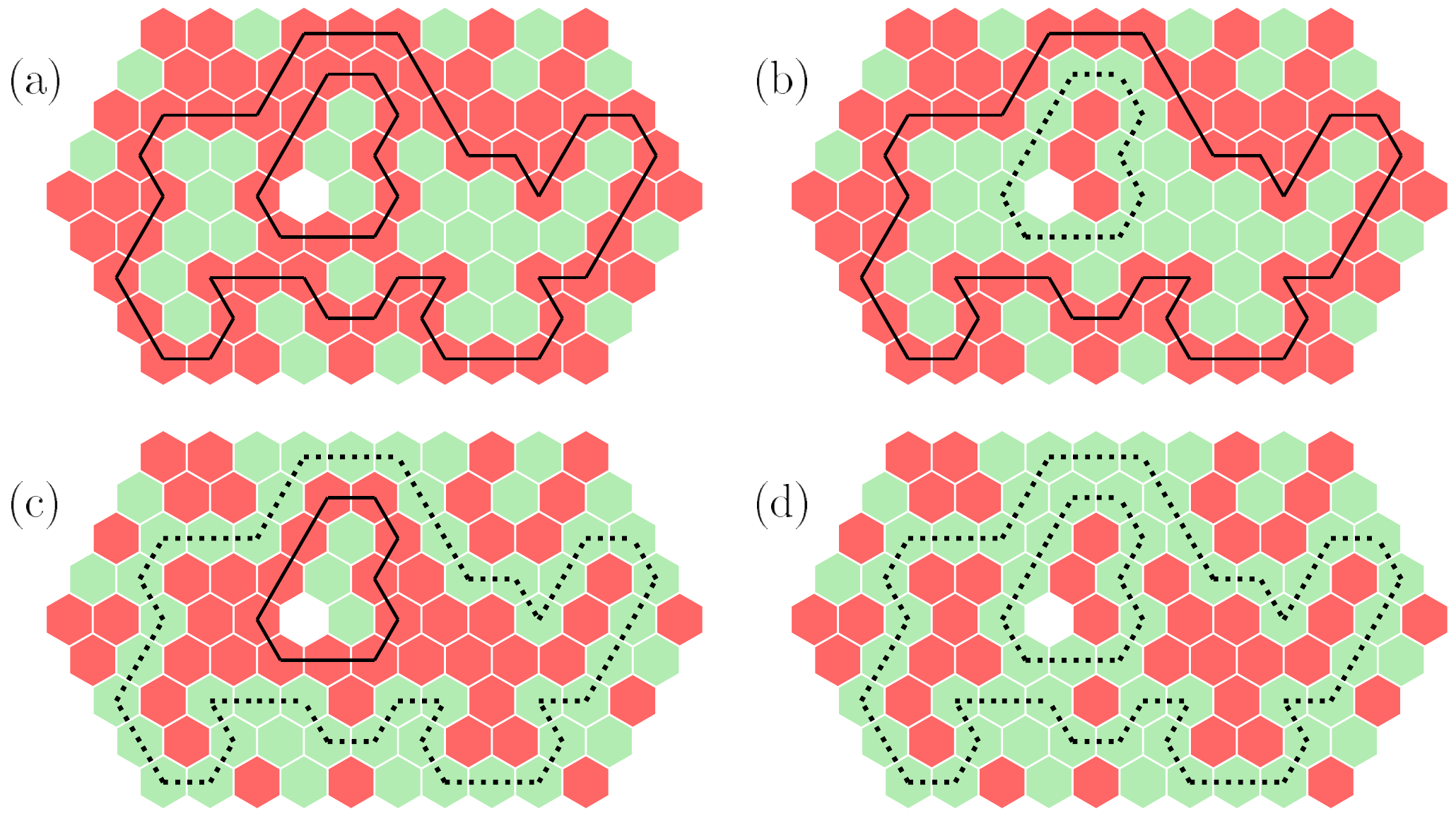}  \vspace{-4mm}
  \caption{Example of a set of $\ell=2$ configurations,
  related by the inversion map $P_i$ in the proof.
  Paths of occupied (empty) sites, denoted by solid (dotted) lines,
  are uniquely located by Algorithm~\ref{alg1}.
  The map $P_1$ leads to (a)$\leftrightarrow$(b) and (c)$\leftrightarrow$(d),
  while $P_2$ gives (a)$\leftrightarrow$(c) and (b)$\leftrightarrow$(d).
  }
  \vspace{-5mm}
  \label{fig:proof}
\end{figure}

Let $\ell$ be the number of these nested paths. 
By $P_i$ we denote the map that inverts (red $\leftrightarrow$ green) the color of path $i$ (counted from the center), and of all sites strictly between path $i$ and path $i-1$ (or between path 1 and the center). 
\Fig{proof} shows four configurations, related to each other by the maps $P_i$.

Since $p_c=\frac12$, the set $\{P_i\}_{i=1}^{\ell}$ generates $2^{\ell}$
equiprobable configurations.
Within this ensemble, ${\cal R} = 1$, if and only if all the paths are occupied.
The entire ensemble can be generated by the $P_i$ from any of its members.
Therefore all configurations are member of precisely one such ensemble, and as a consequence
$W_2 = \langle \scrR \cdot 2^\ell \rangle = 1$.
\quad$\Box$

An essential property used in the proof is that inverting a path of occupied sites 
surrounding the center creates a barrier preventing a path of occupied sites 
to connect the center with the boundary.  This is an obstacle against applying the proof to BSq, where a path of occupied bonds can cross a path of empty bonds, see \cite{SupMat} for more explanation.

The regularity of the lattice and of the domain are not used in the proof, 
but having $p_c = \frac12$ is crucial. However,
$p_c = \frac12$ for any self-matching lattice, 
and in particular for lattices of which all faces are triangles.
Hence, the result is also valid for regular or irregular planar triangulation graphs, 
of any shape and position of the center.

\paragraph{ Discussion ---} We construct a new family of geome\-tric quantities $W_k$ for critical percolation 
in two dimensions and determine a continuous set of critical exponents $\Xnp(k)$ with high precision.
An identity $W_2(L)=1$, independent of the linear size $L$, is found for triangular-lattice 
site percolation and proven for any lattice with only triangular faces.
This implies an exact exponent $\Xnp(2)=0$. 
The universality of the critical exponent $\Xnp$ is well supported by simulations for both bond and site percolation.
Apart from the  special cases $k=2,1,0$, the exact values of $\Xnp$ are unknown.
We conjecture an analytical function, Eq.~(\ref{eq:zeta_conj}),
which reproduces the known exact results and agrees excellently with numerical
estimates of $\Xnp$ for other $k$ values. 
We note that, though Eq.~(\ref{eq:zeta_conj}) is somewhat similar to existing results, proving it remains elusive. 

Future work will involve the $Q$-state Potts model in the Fortuin-Kasteleyn cluster representation, 
which includes bond percolation as a special case for $Q \rightarrow 1$.

 \begin{acknowledgments}
   This work was supported by the National Natural Science Foundation of China (under Grant No.~11625522),
   the Science and Technology Committee of Shanghai (under grant No. 20DZ2210100),
   the National Key R\&D Program of China (under Grant No.~2018YFA0306501),
   and the European Research Council (under the Advanced Grant NuQFT).
 \end{acknowledgments}
\vfill
\bibliographystyle{apsrev4-1}

%
\newpage\begin{widetext}\begin{center}
{\large\bf Supplementary material with \\[3mm]Nested Closed Paths in Two-Dimensional Percolation}\\[3mm]
{Y.-F. Song},
{X.-J. Tan},
{X.-H. Zhang},
{J.L. Jacobsen},
{B. Nienhuis},
{Y. Deng}
\end{center}\vspace{13mm}\end{widetext}
 \section*{Applicability of the proof of $W_2=1$}
The proof that $W_2=1$ for site-percolation on lattices with only
triangular faces fails for bond-percolation on e.g. the square
lattice. This note explains why, and discusses some unsuccesful attempts to remedy this.

The first issue is that in the proof we consider paths both on occupied
and on empty sites, occupied paths and empty paths, for short.  For bond percolation there is a choice how to construct the empty paths; this is between
(i) the bonds of the lattice that happen to be empty, or
(ii) the bonds of the dual lattice at the positions where the primal bond is empty.
The proof uses two properties of the relation between occupied and empty paths: (a) an occupied path becomes an empty path (and v.v.) under inversion, and (b) an empty path and an occupied path cannot cross.  With choice (i) we have property (a) but not (b), and with choice (ii) we property (b) but not (a).  \Fig{dual} below illustrates these facts.
As a consequence, a straightforward translation of the proof for BSq is not possible.
\begin{figure}[hb]
    \setlength{\unitlength}{8mm}\begin{picture}(11,3)(0,-.2)
    \put(.5,0){\includegraphics[width=10\unitlength]{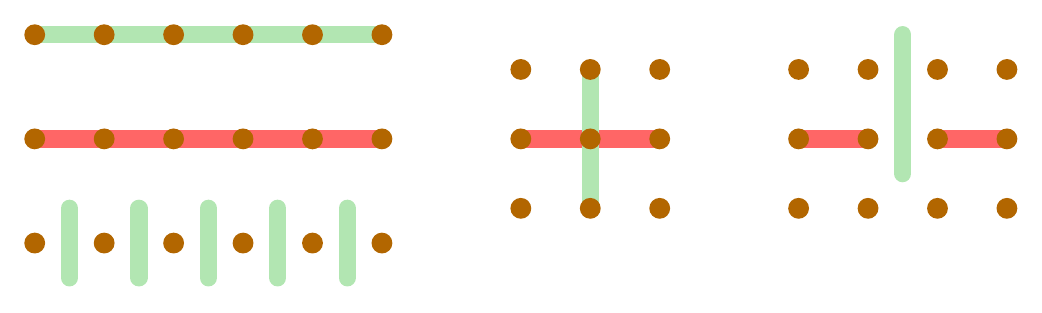}}
    \put(.1,.5){\large c}  \put(.1,1.5){\large b}  \put(.1,2.5){\large a}
   \put(6.1,.3){\large d}\put(9,.3){\large e} 
  \end{picture}
\caption{Properties of the inversion map.  Occupied bonds are marked red, and empty bonds green.  A path of occupied bonds (b), converted by an inversion transformation, to  (a) an empty path on the same lattice, under choice (i),  or under choice (ii) to (b) a sequence of dual bonds, not forming a path.  Part (d) shows how an occupied path and an empty path can cross, but (e) an occupied path cannot cross an empty path on the dual lattice. \label{fig:dual}} \end{figure}
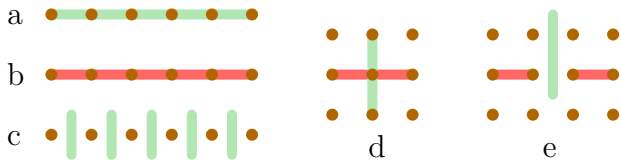

One may consider an alternative definition of $W_k$ that would allow to reconstruct the proof.
A central concept in the proof is the inversion transformation $P_i$ of a path and (a part of) its interior. Under such transformation a path remains a path but changes color.  This seems to make choice (ii) hopeless.

So we concentrate on choice (i), and consider bonds and paths on the primal lattice only.  Then we have to deal with the difficulty that paths can cross each other.  We rule out crossing paths of the same color by construction, so that it is still possible to define innermost paths.  But crossing paths of different colors are difficult to rule out, as in this case one must give precedence to one or the other.

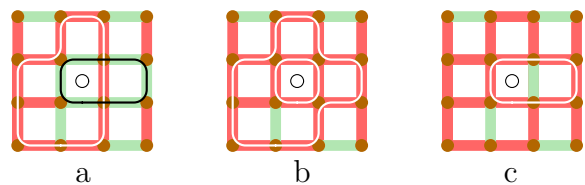
\begin{figure}[h]\setlength{\unitlength}{1cm}\begin{picture}(8,3)
  \put(.05,.2){\includegraphics[width=8cm]{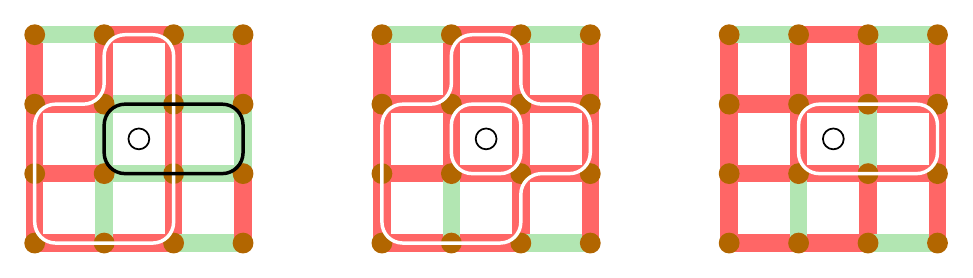}}
  \put(1.1,0){\large a}
  \put(4,0){\large b}
  \put(6.8,0){\large c}
\end{picture}
\caption{Attempts to define the inversion maps. Part (a) shows a configuration with both an occupied path (marked white) and an empty path (marked black) surrounding the center (black circle).  After the empty path in inverted, in (b), there are two occupied paths as desired, but not (necessarily) at the same location as the original ones. If, besides the empty path also its interior is inverted (c), there is only one (indepedent) occupied path left.\label{fig:inversions}}
\end{figure}

Thus the restriction that the center is connected to the boundary over
an occupied path, does not rule out the existence of an empty path
surrounding the center.   It is tempting to redefine $W_k$ by giving the
weight $k$ to each occupied path surrounding the center, while demanding
there are no empty paths surrounding the center, this condition
replacing the one that an occupied path connects the center  to the boundary.  But for a configuration with occupied and empty paths, both surrounding the center, and crossing each other, we did not succeed to define the transformations $P_i$ appropriately.  The difficulties are illustrated in \Fig{inversions}a showing a configuration in which the center (black circle) is surrounded by both an occupied and an empty path.  If the empty path is inverted, the resulting configuration has two occupied paths, but with a different structure (b).  One may argue that in the original definition, not only the path itself, but also (part of) its interior was inverted.  But inverting the empty path with its interior results in an even less hopeful configuration, with only one surrounding path left (\Fig{inversions}c).

In conclusion, we have not found a variant definition of $W_k$ such that we can prove that $W_2=1$ also for BSq.
\end{document}